# High efficiency and low absorption Fresnel compound zone plates for hard X-ray focusing

Kuyumchyan A.*[a], Isoyan A.*[a], Shulakov E.[a], Aristov V.[a], Kondratenkov M.[a], Snigirev A.[b], Snigireva I.[b], Souvorov A.[c], Tamasaku K.[c], Yabashi M.[c], Ishikawa T.[c], Trouni K.[d]

[a]Institute of Microelectronics Technology and High Purity Materials of Russian Academy of Science; [b]European Synchrotron Radiation Facility, B.P. 202, F - 38043 Grenoble, France; [c]SPring-8, JASRI, 1-1-1 Kouto, Mikazuki-cho, Sayo-gun Hyogo 679 5198, Japan; [d]International Academy of Science and Technology, P.O. Box 4385, CA 91222, USA

**ABSTRACT**

Circular and linear zone plates have been fabricated on the surface of silicon crystals for the energy of 8 keV by electron beam lithography and deep ion plasma etching methods. Various variants of compound zone plates with first, second, third diffraction orders have been made. The zone relief height is about 10 mkm, the outermost zone width of the zone plate is 0.4 mkm. The experimental testing of the zone plates has been conducted on SPring-8 and ESRF synchrotron radiation sources. A focused spot size and diffraction efficiency measured by knife-edge scanning are accordingly 0.5 mkm and 39% for the first order circular zone plate.

## 1. INTRODUCTION

Many types of optical devices, including curved crystals and mirrors [1-5], capillaries and waveguides [6,7], Fresnel zone plates [8], Bragg-Fresnel zone plates [9], compound zone plates [10,11], kinoform lenses [12,13], grazing incidence phase zone plates [14], parabolic lenses [15] are used to focus X-ray beams.
Axial X-ray optical elements attract great interest lately. It is connected with their practical application in phase contrast analysis of different substances [16] in X-ray fluorescence microtomography [17], and others.
Linear resolution and diffraction efficiency defined by their fabrication technology are important features of X-ray optical elements. The main difficulty of fabricating zone plates with Fresnel surface relief for high energy is deep etching of zone structures with alternate aspect ratio in the single process.
The process of fabrication zone plates is described in this paper. The results of the fabricated zone plates testing on the SPring-8 synchrotron radiation source is given.

## 2. THE FABRICATION OF COMPOUND ZONE PLATES

The boundaries of half-zones for a compound zone plates, working in $m$-order of diffraction are defined as:

$$r_{mn} = \begin{cases} \sigma(mn-j)^{1/2}, & n-odd \\ \sigma(mn)^{1/2}, & n-even \end{cases} \quad (1)$$

where $\sigma=(\lambda f)^{1/2}$, $\lambda$ - X-ray radiation wavelength, $f$ – a focus distance, $m$-order of diffraction, $n$-the number of half zones, $j$ is connected with slitness $S$ by the relation $S=(m-j)/2m$. The sizes of even and odd halh zones are represented by:

$$\Delta r_{mn} = \begin{cases} \sigma(m-j)/2(mn)^{1/2}, & n-odd \\ \sigma(m+j)/2(mn)^{1/2}, & n-even \end{cases} \quad (2)$$

if $n \gg 1$ the sizes of neighbouring zones are related as $(m-j)/(m+j)$.
If $S=1/2$ ($j=0$) even diffraction orders are absent. The diffraction efficiency of different orders can be modified by changing the slitness. In what follows we shall consider the absorption absent and the zone plate under study phase. The criterion of selection on the basis of the parameter j for achieving maximal efficiency $\varepsilon=4/\pi^2 m^2$ in $m$ – order of diffraction is given by

$j$ – a whole number, $|j| < m$, $j$ – an even number for odd $m$ and an odd number for even $m$. (3)

The intensity of undiffracted radiation (zero diffraction order) for the component with $m$ – order is defined a $\varepsilon_0=(j/m)^2$.

The change of the zone contribution to the intensity during the slitness variation is shown in Fig.1.

The usage of external component with slitness different from 1/2 enables to choose any diffraction orders in the, both odd and even orders, in the compound zone plate with maximal efficiency in according with condition (3).

The components of the compound zone plate will create smaller background in the region of the main focus, as the main part of the background of higher diffraction orders will be distributed to the zero order which is screened by the inner component acting in the first diffraction order.

From the point of view of diffraction the structures described by the parameters $+j$ and $-j$ are equivalent. At the same time the neighbouring half zones widths differ substantially. It permits to widen the critical half zone and to enlarge the aperture of the outermost component to $A_m=A_1(m+|j|)$, where $A_1$ and $A_m$ – are external apertures of the first order component and m diffraction order.

The height of the zone $h$ at which the phase is changed to $\pi$ is defined as $h=\lambda/\chi_{Or}$, where $\chi_{Or}$ Fuhre coefficient of the polarized crystal real part decomposition. The relief height for silicon with radiation energy of 8.05 keV is 10.18 mkm.

Zone plates have been made of 450 mkm thick silicon, orientation (111) and the plate size 15x20 mm. The membrane was 15-20 mkm thick to reduce absorption of X-ray radiation (Fig.2).

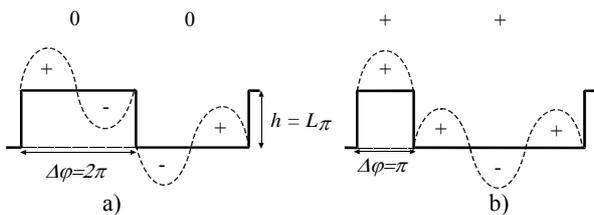
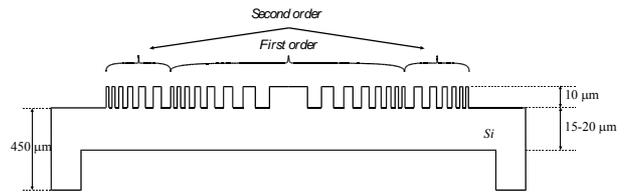

Fig.1: The contribution to the intensity of main focus of the second order zones for structures with different relative slitness. a) S=1/2, b) S=1/4.

Fig.2: The structure of the compound zone plate.

The process of fabrication zone plates is shown in Fig.3. 10 mkm thick photo resist is applied to the back side of the silicon. Then a 15-20 mkm thick membrane is formed on the silicon substrate by photolithography and isotrope and anisotrope wet etching. Then electron resist 300 nm thick is applied to the processed surface. The structures of zone plate is formed of the 200 nm thick nickel on the membrane by electron beam lithography and lift-off technology. Then the zone plate structures is etched by ion plasma etching, the depth of etching is about 10 mkm. Different types of compound zone plates for hard X-ray radiation have been made (Fig.4.).

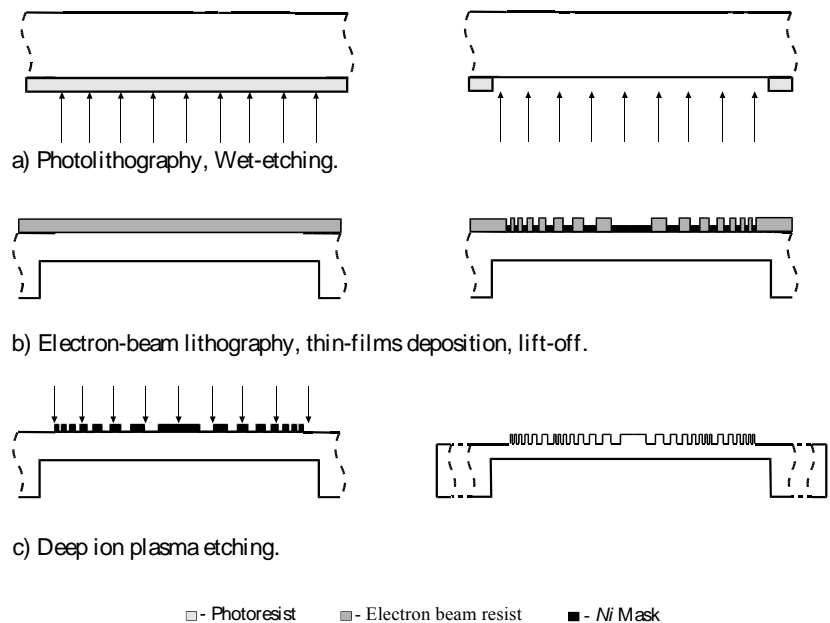

Fig.3: The process of the zone plate fabrication.

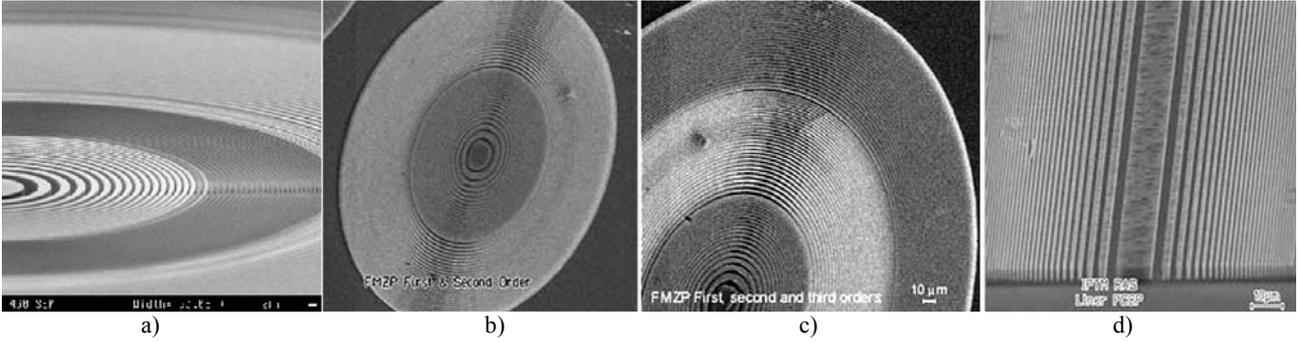

Fig.4: SEM image of the fabricated compound Fresnel zone plates. a) a compound zone plate of the first and third order, b) a compound zone plate of the first and second order, c) a compound zone plate of the first, second and third order, d) optimized linear zone plate of the first and third order.

## 3. EXPERIMENTAL TESTING OF ZONE PLATES

The circular zone plate of the first order has been tested on the synchrotron radiation source Spring-8. The experiment has been conducted at the distance of 1 kilometer of the station BL29XU [18]. The energy E is 8 keV. Cryogen cooled double crystal monochromator has been used. The distance between source and the experimental set-up is about 1 km. The vertical source size is 25 mkm (FWHM). A schematic diagram of the experimental set-up is shown in Fig.5. The entrance of slit collimator corresponds to the aperture of zone plate. An Ion chamber is used to monitor the incident radiation. The image of source with the reduction of approximately of 2000 in registered in the focal plane of the zone plate. Thus, the image reduction, related to the size of the source, is 0.01 mkm, and, therefore, the function of spot broadening is registered in the focal plane. The width of the slit analyzer is 5 mkm in the horizontal direction.

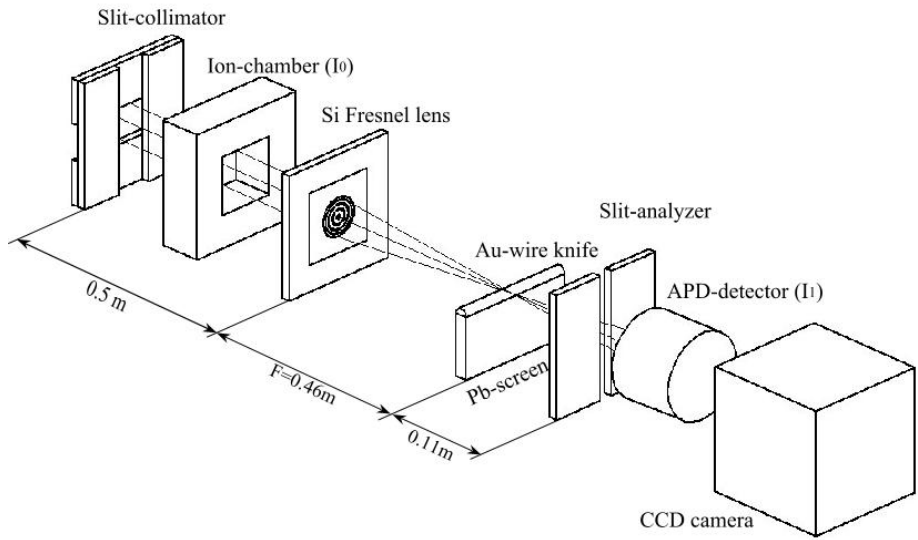

Fig.5: A schematic diagram of the experimental set-up.

The parameters of the tested zone plate: the first zone radius is 8.42 mkm, the number of zones is 112, the outermost zone width is 0.4 mkm, aperture is 178.2 mkm, focal distance is 46 cm, the relief height is 10.5 mkm, phase shift for neighbouring half zones is $1.037\pi$, the thickness of the membrane is 16 mkm. The coefficient of the membrane transmittance for the energy 8 keV is 79%. The average coefficient of the surface relief transmittance is 93 %.

The focal spot size is measured by a conventional knife-edge scanning method. The distribution of the intensity during the knife-edge scanning in the focal plane of the zone plate is shown in Fig.6a. Part of knife-edge scanning in the focal plane and its derivation is represented in Fig.6b.

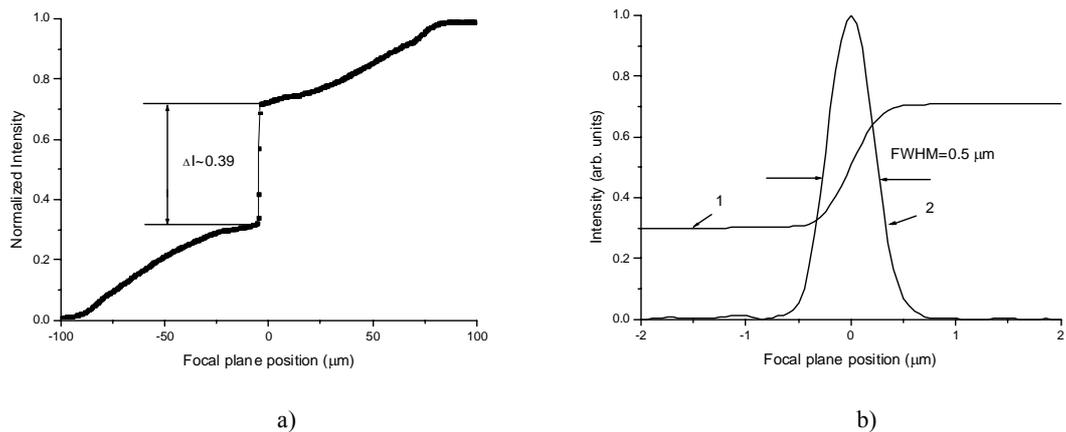

a)                                                        b)

Fig.6: The profiles of intensity distribution at knife-edge scanning. a) knife-edge scanning of the zone plate aperture, b) part of knife-edge scanning in the focal region (1) and its derivative (2).

The focal width at half maximum (FWHM) is 0.5 mkm. This value agrees well with the theoretical evaluation of the focus size 0.488 mkm. the disagreement between the theory and the experiment is mainly connected with error of fabricating outermost zones. The diffraction efficiency of the first order zone plate, measured by knife-edge scanning is about 39 %.
An overexposed image of the zone plate formed by undiffracted radiation in the focal plane is shown in Fig.7.

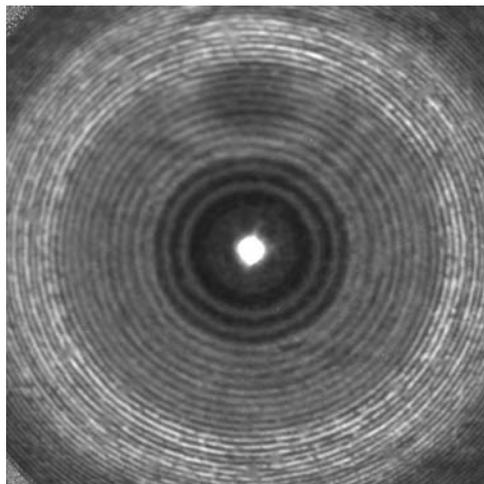

Fig.7: A topographic scheme of the zone plate in the focal plane. The image size is 762x762 pixels, where one pixel equal 0.2 mkm.

## 4. CONCLUSIONS

High efficiency and low absorption compound Fresnel zone plates with the focal length from 0.5 m to 1 m, minimal outermost zone width of 0.4 mkm and aspect ratio up to 25 have been fabricated by electron beam lithography and deep ion plasma etching of silicon. The experimental testing of the first order diffraction zone plates has been conducted on the synchrotron radiation source Spring-8. The measured focal size of the zone plate is 0.5 mkm, experimental value of relative diffraction efficiency measured by knife-edge scanning is 39 %. It means that zone plates produced by us are close to pure phase ones. The resolution and efficiency of zone plates agree well with theoretical evaluation.
The testing of the compound zone plates on synchrotron radiation sources will be carried on in the near future.

## ACKNOWLEDGEMENTS

The authors are grateful to Mr. S. Pyatkin, Mr. V. Yunkin, Mr. I. Amirov for the assistance in fabricating zone plates. The work has been carried on with financial support of RFBR № 01-02-16472, RFBR № 02-02-22009.

* Arkuyumchyan@mtu-net.ru; aisoyan@mail.ru phone +7 09652 4-40-81; fax +7 095 962-80-47; http://www.ipmt-hpm.ac.ru; Institute of Microelectronics Technology and High Purity Materials of Russian Academy of Science,142432, Russia, Chernogolovka